\documentclass[twocolumn,showpacs,amssymb,aps]{revtex4}
\usepackage{graphicx}

\begin{document}

\title{Electron-hole pair condensation in graphene bilayer}

\author{Yu.\,E. Lozovik}
\email{lozovik@isan.troitsk.ru}
\author{A.\,A. Sokolik}

\affiliation{Institute of Spectroscopy, Russian Academy of Sciences, 142190 Troitsk, Moscow Region, Russia}

\begin{abstract}
We consider the pairing of electrons and holes due to their Coulomb attraction in two parallel, independently gated
graphene layers, separated by a barrier. At weak coupling, there exist the BCS-like pair-condensed state. Despite the
fact that electrons and holes behave like massless Dirac fermions, the problem of BCS-like electron-hole pairing in
graphene bilayer turns out to be rather similar to that in usual coupled semiconductor quantum wells. The distinctions
are due to Berry phase of electronic wave functions and different screening properties. We estimate values of the gap
in one-particle excitation spectrum for different interlayer distances and carrier concentrations. Influence of
disorder is discussed. At large enough dielectric susceptibility of surrounding medium, the weak coupling regime holds
even at arbitrarily small carrier concentrations. Localized electron-hole pairs are absent in graphene, thus the
behavior of the system \textit{versus} coupling strength is cardinally different from usual BCS-BEC crossover.
\end{abstract}


\maketitle

Recent progress in experimental technology allowed a fabrication of graphene, one atomic layer separated from graphite
crystal [1--4]. A lot of theoretical attention to graphene has emerged due to peculiar properties of its band
structure, consisting in linear dispersion of electron energy near two inequivalent points of the Brillouin zone
[5--7]. An electron wave function close to these points is well described by the two-dimensional Dirac equation for
massless particles [6,~7] with the Fermi velocity $v_\mathrm{F}\approx10^6\,\mbox{m/s}\approx c/300$ playing the role
of effective ``speed of light" [2]. Several peculiar transport phenomena in graphene has been discovered
experimentally, e.g., anomalous quantum Hall effect [3] and minimal conductivity [2,~8]. Unique properties of graphene,
such as unusually high mobility of charge carriers [1] and a phase coherent transport [9,~10], allow to propose it as a
base of future nanoelectronic devices [11--13].

In the present work we consider the formation of the condensate of spatially separated electron-hole pairs in bilayer
graphene structure when interlayer tunneling is negligible. A condensation and superfluidity of spatially separated
electron-hole pairs in usual coupled semiconductor quantum wells (CQW) due to their Coulomb attraction has been
proposed theoretically in [14]. A nondissipative motion of resulting pairs leads to appearance of persistent electric
currents, flowing in two layers in opposite directions (contrary to 3D case, where phase fixation leads to a formation
of excitonic insulator state [15]). In the present Letter we consider another physical realization of a two-dimensional
electron-hole system
--- graphene bilayer. Its schematic setup is shown on Fig.1.

Two parallel graphene sheets are separated by a dielectric layer of thickness $D$, large enough to neglect tunneling
between them. By applying the gate voltage $V_\mathrm{g}$ between graphene sheet and a gate electrode, isolated from it
by dielectric layer, one can adjust the charge carrier concentration, fixing the chemical potential $\mu$ at any
desired level [1] (in addition to electrical one, chemical or electrochemical [16] doping of graphene is also
possible). Chemical potentials in either of graphene layers may be adjusted independently. We consider the case of
equal densities, when in the top layer the chemical potential is $\mu>0$, and charge carriers are electrons, whereas in
the bottom layer the chemical potential is $-\mu<0$, and charge carriers are holes. At weak coupling conditions, the
system is unstable with respect to Bardeen-Cooper-Schrieffer (BCS) interlayer pairing of electrons and holes due to
their Coulomb attraction.

\begin{figure}[b]
\begin{center}
\includegraphics[width=0.9\columnwidth]{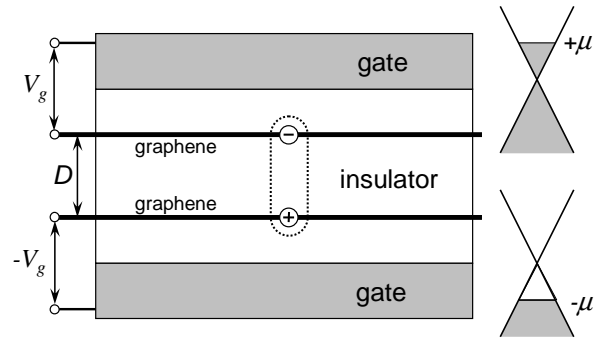}
\caption{Schematic set-up of a system for realization of a pairing of spatially separated electron and holes in
graphene bilayer. In the right: chemical potential positions in two graphene layers, adjusted by gate voltages
$V_\mathrm{g}$ and $-V_\mathrm{g}$.}
\end{center}
\end{figure}

Consider the effective Hamiltonian of the system, responsible for pairing of electrons from the top graphene layer and
holes from the bottom layer. The influence of remaining part of the total Hamiltonian, corresponding to electron and
hole interactions within individual graphene layers, manifests itself via screening of interlayer Coulomb interaction.
The effective Hamiltonian can be presented in the form:
\begin{eqnarray}
H_0=\sum_{\mathbf{k}}\xi_\mathbf{k}(a_{\mathbf{k}}^+a_{\mathbf{k}}+
b_{\mathbf{k}}^+b_{\mathbf{k}})+\frac1S\times\nonumber\\
\times\sum_{\mathbf{k}_1,\mathbf{k}_2,\mathbf{q}} V(\mathbf{q})\cos\frac{\varphi_1}2\cos\frac{\varphi_2}2
a_{\mathbf{k}_1+\mathbf{q}}^+b_{\mathbf{k}_2-\mathbf{q}}^+b_{\mathbf{k}_2}a_{\mathbf{k}_1},\label{ham1}
\end{eqnarray}
where $a_{\mathbf{k}}$ and $b_{\mathbf{k}}$ are destruction operators for Dirac electron and hole quasiparticles with in-plane momentum $\mathbf{k}$,
$\xi_{\mathbf{k}}=\hbar v_\mathrm{F}|\mathbf{k}|-\mu$ is a quasiparticle energy measured from the Fermi level, $V(\mathbf{q})$ is the potential of
screened electron-hole interaction, $\mathbf{q}$ is a momentum transmitted via Coulomb scattering, $\varphi_1$ and $\varphi_2$ are scattering angles
for electron and hole, and $S$ is the area of a bilayer. The graphene-specific factor $\cos(\varphi_1/2)\cos(\varphi_2/2)$ results after folding over
components of a spinor electron wave function. The pairing is completely degenerate on spin and valley quantum numbers of electron and hole and thus
we ignore these degrees of freedom.

Within the framework of BCS approach for a pairing of electrons and holes with opposite momenta [17], the Hamiltonian
(\ref{ham1}) transforms into the form:
\begin{eqnarray}
H=\sum_{\mathbf{k}}\xi_\mathbf{k}(a_{\mathbf{k}}^+a_{\mathbf{k}}+
b_{\mathbf{k}}^+b_{\mathbf{k}})+\frac1S\times\nonumber\\
\times\sum_{\mathbf{k},\mathbf{q}} V(\mathbf{q})\frac{1+\cos\varphi}2
a_{\mathbf{k}+\mathbf{q}}^+b_{-\mathbf{k}-\mathbf{q}}^+b_{-\mathbf{k}}a_{\mathbf{k}},\label{ham2}
\end{eqnarray}
where $\varphi$ is angle between $\mathbf{k}$ and $\mathbf{k}+\mathbf{q}$, i.e. scattering angle, equal for electron
and hole. Here the factor $(1+\cos\varphi)/2$ means an overlap of initial and final electron states. It originates from
Berry phase of electronic wave functions in graphene and have no analogue in CQW [18].

The Hamiltonian (\ref{ham2}) can be diagonalized by Bogolyubov transformation:
\begin{eqnarray}
a_{\mathbf{k}}=u_{\mathbf{k}}\alpha_{\mathbf{k}}+v_{\mathbf{k}}\beta_{-\mathbf{k}}^+,\quad
b_{-\mathbf{k}}=u_{\mathbf{k}}\beta_{-\mathbf{k}}-v_{\mathbf{k}}\alpha_{\mathbf{k}}^+.\nonumber
\end{eqnarray}
Introducing usual notations
\begin{eqnarray}
u_{\mathbf{k}}^2=\frac12\left(1+\frac{\xi_{\mathbf{k}}}{E_{\mathbf{k}}}\right),\quad
v_{\mathbf{k}}^2=\frac12\left(1-\frac{\xi_{\mathbf{k}}}{E_{\mathbf{k}}}\right),\nonumber\\
u_{\mathbf{k}}v_{\mathbf{k}}=\frac12\frac{\Delta_{\mathbf{k}}}{E_{\mathbf{k}}},\quad
E_{\mathbf{k}}=\left(\xi_{\mathbf{k}}^2+\Delta_{\mathbf{k}}^2\right)^{1/2},\nonumber
\end{eqnarray}
we derive the self-consistent gap equation
\begin{eqnarray}
\Delta_{\mathbf{k}}=-\frac1{4\pi^2}\int d\mathbf{q}V(\mathbf{q})\frac{1+\cos\varphi}2
\frac{\Delta_{\mathbf{k}+\mathbf{q}}}{2E_{\mathbf{k}+\mathbf{q}}},\label{gap_eq1}
\end{eqnarray}
where we turned from summation to integration on $\mathbf{q}$. Note that Eq.\,(\ref{gap_eq1}) differs from the analogous one for CQW in two aspects.
The first one is the linear dependence of $\xi_\mathbf{k}$ on $\mathbf{k}$ instead of a quadratic one for CQW; however, near the Fermi energy it can
be linearized, hence this difference is not essential. The second aspect is the presence of the overlap factor $(1+\cos\varphi)/2$, which suppresses
backscattering in graphene. At weak coupling this factor turns into unity due to a predominance of small scattering angles in the integral in
(\ref{gap_eq1}), but at stronger coupling it can weaken the pairing.

The main contribution to the integral in (\ref{gap_eq1}) comes from the region near to the Fermi energy, where the dynamically screened interlayer
electron-hole interaction $V(\mathbf{q},\omega)$ is attractive. In random phase approximation,
\begin{eqnarray}
V(\mathbf{q},\omega)=\frac{-v_{\mathbf{q}}e^{-qD}}{1-v_{\mathbf{q}}(\chi_1+\chi_2)+
v_{\mathbf{q}}^2\chi_1\chi_2(1-e^{-2qD})},
\label{V_c}
\end{eqnarray}
where $v_{\mathbf{q}}=2\pi e^2/\varepsilon q$ is the bare Coulomb interaction, $\varepsilon$ is the dielectric constant
of a surrounding medium, $\chi_1$ and $\chi_2$ are dynamic polarizabilities within the top and bottom graphene sheets.
In the case of equal densities, due to the particle-hole symmetry, polarizabilities are equal in both graphene sheets:
$\chi_1=\chi_2=\chi$. The equation
\begin{eqnarray}
1-2v_{\mathbf{q}}\chi(\mathbf{q},\omega)+v_{\mathbf{q}}^2\chi^2(\mathbf{q},\omega)(1-e^{-2qD})=0\label{plasm}
\end{eqnarray}
describes two branches $\omega_{\pm}(q)$ of a plasmon dispersion in the system, corresponding to in-phase and antiphase
plasma oscillations [14,~19,~20].

The weak coupling (or BCS) regime takes place when the region of pairing near to the Fermi energy is narrow with
respect to the Fermi energy itself. In such a case, the radial integration on $\xi\equiv\xi_{\mathbf{k}+\mathbf{q}}$ in
the gap equation (\ref{gap_eq1}) can be decoupled from the integration on the polar angle $\varphi$. We restrict the
integration on $\xi$ to the region limited by the cutoff energy $\hbar\tilde\omega$, where the \emph{dynamically}
screened interaction (\ref{V_c}) is attractive. As for the integration on $\varphi$, we perform it with the
\emph{statically} screened electron-hole attractive potential $V(\mathbf{q})\equiv V(\mathbf{q},0)$.

Assume that the main part of the integral in (\ref{gap_eq1}) comes due to integrands with small $q$, not larger than
$\tilde{q}$ by order of magnitude. Let define $\tilde{q}$ as satisfying the equation $V(\tilde{q})=V(0)/2$. The
explicit form of $V(\mathbf{q})$ is determined by the static polarizability [20--22]
\begin{eqnarray}
\chi(\mathbf{q},0)=-\frac{g_\mathrm{s}g_\mathrm{v}\mu}{2\pi\hbar^2v_\mathrm{F}^2}\equiv-\frac\varepsilon{\pi
e^2a},\label{chi_TF}
\end{eqnarray}
where $a$ is the Thomas-Fermi screening length in graphene. Coefficients $g_\mathrm{s}=g_\mathrm{v}=2$ arise due to spin and valley degeneracy of
electron states in graphene.

Let introduce the Fermi momentum $k_0=\mu/\hbar v_\mathrm{F}$ and the dimensionless parameter
\begin{eqnarray}
\alpha=\frac{2\varepsilon\hbar
v_\mathrm{F}}{g_\mathrm{s}g_\mathrm{v}e^2}=\frac1{2r_\mathrm{s}}\approx0.23\times\varepsilon,\nonumber
\end{eqnarray}
where the dimensionless Wigner-Seitz radius $r_\mathrm{s}$ measures the ratio of the characteristic Coulomb energy of
quantum system to its characteristic kinetic energy [20]. In 2D semiconductor system $r_\mathrm{s}$ increases with
decreasing carrier density, but in graphene it is determined only by a dielectric constant $\varepsilon$ of surrounding
medium.

There are three characteristic distances in the system, namely $a$, $D$ and the mean separation between charge carriers
within each graphene layer $l\sim1/k_0$. A behavior of the system depends on a relation between $a$, $D$ and $l$. As
can be shown, $a/l=\alpha$ and $D/l=k_0D$. Therefore, a behavior of the system is governed by two dimensionless
parameters, $\alpha$ and $k_0D$.

The characteristic momentum $\tilde{q}$ can easily be found from (\ref{V_c}) and (\ref{chi_TF}) in two limiting cases:
$\alpha\ll k_0D$ and $\alpha\gg k_0D$. In the first one, an effective momentum cutoff occurs due to the factor
$\exp(-qD)$ in the numerator of (\ref{V_c}), thus we get $\tilde{q}\approx2/D$. In the second one, it is determined by
the Thomas-Fermi screening, and we get $\tilde{q}\approx4k_0/\alpha$. Both results can be written as
\begin{eqnarray}
\tilde{q}=\min\left(\frac{4k_0}\alpha,\frac2D\right)\label{q_tilde}.
\end{eqnarray}

The cutoff energy $\hbar\tilde\omega$ is determined by a characteristic frequency of the lower branch of plasma
oscillations and can be estimated as $\hbar\tilde\omega=\hbar\omega_-(\tilde{q})$. To the first order in the
electron-electron interaction, the dynamic polarizability at $q\rightarrow0$ and $\omega>v_\mathrm{F}q$ is [20,~22]
\begin{eqnarray}
\chi(\mathbf{q},\omega)=\frac{g_\mathrm{s}g_\mathrm{v}\mu q^2}{4\pi\hbar^2\omega^2}.\label{chi_dyn}
\end{eqnarray}
In the case $\alpha\ll k_0D$, from (\ref{plasm}) and (\ref{chi_dyn}) we find
$\omega_-(q)=v_\mathrm{F}q(k_0D/\alpha)^{1/2}$ and $\omega_+(q)=v_\mathrm{F}(2k_0q/\alpha)^{1/2}$. Thus the cutoff
energy is $\hbar\tilde\omega=2\mu/(k_0D\alpha)^{1/2}$. In the case $\alpha\gg k_0D$, the approximate expression
(\ref{chi_dyn}) is inapplicable, because the lower branch of a plasmon dispersion, formally found with it, falls into a
single-particle excitation continuum $\omega<v_\mathrm{F}q$. Actually, in this case $\omega_-(q)=v_\mathrm{F}q$ and
$\omega_+(q)=v_\mathrm{F}(2k_0q/\alpha)^{1/2}$, and the cutoff energy is $\hbar\tilde\omega=4\mu/\alpha$.

Electron-hole Cooper pairs have a size of order of $1/\tilde{q}$ in the in-plane direction. The weak coupling requires
a large pair size relative to mean separation between nearest pairs, i.e. $\tilde{q}l\ll1$. Using (\ref{q_tilde}), we
conclude, that the weak coupling regime occurs when at least one of parameters $\alpha$ or $k_0D$ is large with respect
to unity.

It is not of necessity in our case for the pairing to be $s$-wave. Due to spatial separation, Pauli exclusion principle does not impose any
conditions on relative angular momentum as well as spins and valleys of paired particles. To seek an $l$-wave solution of (\ref{gap_eq1}) we assume
that $\Delta_\mathbf{k}=\Delta\exp(il\varphi_\mathbf{k})$ at $|\xi|\leqslant\hbar\tilde\omega$ and $\Delta_\mathbf{k}=0$ at
$|\xi|>\hbar\tilde\omega$. The approximate solution is
\begin{eqnarray}
\Delta=2\hbar\tilde\omega\exp\left\{-\frac{4\pi\alpha}{V_l}\right\}.\label{gap_int}
\end{eqnarray}
Here the dimensionless $l$-wave harmonic of $V(\mathbf{k})$ (corrected by the Berry phase factor) is
\begin{eqnarray}
V_l=\int\limits_0^{2\pi}\frac{e^{-k_0Dx}(1+\cos\varphi)/2}{x+4/\alpha+4(1-e^{-2k_0Dx})/\alpha^2x}
\:e^{-il\varphi}d\varphi,
\end{eqnarray}
and $x=q/k_0=2\sin(\varphi/2)$ in the integral.

The condensation will lead to the $l$-wave pairing with the largest $\Delta$ and thus the largest $V_l$. We have found
numerically that at any values of $\alpha$ and $k_0D$ the $s$-wave pairing dominates, though at very weak coupling a
difference between $V_l$ for different $l$ becomes negligible. So we consider only $s$-wave pairing hereafter. Deriving
asymptotic expressions for $V_l$ and substituting them into (\ref{gap_int}) we find the gap for different relations
between $\alpha$, $k_0D$ and unity.

At $\alpha\ll k_0D$ we get
\begin{eqnarray}
\Delta=\frac{4\mu}{(k_0D\alpha)^{1/2}}\exp\left\{-8\pi k_0D\left(1+\frac{k_0D}\alpha\right)\right\}.\label{gap1}
\end{eqnarray}
This expression is suitable for values of $\alpha$, both small and large with respect to unity.

At $\alpha\gg k_0D$ we distinguish cases of small and intermediate interlayer distances $D$. In the first case,
$k_0D\ll1\ll\alpha$, the gap is
\begin{eqnarray}
\Delta=\frac{8\mu}\alpha\exp\left\{-\frac{2\pi\alpha}{\ln(1+\alpha/2)}\right\}.\label{gap2}
\end{eqnarray}
In the second case, $1\ll k_0D\ll\alpha$, the gap is
\begin{eqnarray}
\Delta=\frac{8\mu}\alpha\exp\left\{-\frac{2\pi\alpha}{\ln(\alpha/4k_0D)-\gamma}\right\},\label{gap3}
\end{eqnarray}
where $\gamma\approx0.577$ is the Euler constant.

We have considered the weak coupling regime, but what can occur at stronger coupling? In CQW, at $T=0$, on increase of
a coupling strength there exist a crossover from BCS-like state to Bose-Einstein condensation (BEC) in a dilute gas of
localized, non-overlapping electron-hole pairs (or excitons for quasi-equilibrium state created after laser pumping).
At ${T\neq0}$, both BCS-like state and gas of local pairs are in superfluid state below the temperature of
Kosterlitz-Thouless transition to the normal state [14]. In graphene, there are no bound solutions for the Dirac
problem of single electron in attractive potential due to the absence of a gap in the energy spectrum. Similarly, as
can be shown, there are no localized electron hole-pairs in a bilayer. Therefore a behavior of graphene electron-hole
bilayer on increase of the coupling strength will be cardinally different from BCS-BEC crossover in CQW. Strong
coupling regime will be studied in subsequent publication. Note, that in a perpendicular magnetic field the existence
of localized magnetoexcitons makes BCS-BEC crossover possible [23,~24].

Consider the weak coupling conditions more closely at $\alpha\ll1$ or $\alpha\sim1$. Such values of $\alpha$ can be
realized with, e.g., commonly used $\mathrm{SiO}_2$ substrate ($\varepsilon\approx4$). In this case the coupling
strength is determined only by a value of $k_0D$. The Fermi momentum $k_0$ is proportional to $\mu$, which can be tuned
from zero to maximal values of $\approx0.3\,\mbox{eV}$ in electrically doped graphene [1,~2]. The weak coupling regime
($k_0D\gg1$) can be achieved with reasonable carrier concentrations at any interlayer distance $D>100\,\mbox{\AA}$. On
the other side, by tending $\mu$ to zero, one can always achieve a strong coupling regime. Thus, the whole transition
from weak to strong coupling can be realized experimentally by changing the gate voltage.

The case $\alpha\gg1$ takes place at large values of the dielectric constant $\varepsilon$ of surrounding medium (at least $\varepsilon>5$), which
can be achieved with, e.g., $\mathrm{HfO}_2$ ($\varepsilon\approx25$). In this case, the weak coupling regime sustains even at $\mu\rightarrow0$,
i.e. at arbitrarily small carrier concentrations, and the gap tends to zero as $\Delta\propto\mu$ according to (\ref{gap2}). This provides a
remarkable contrast with CWQ, where a strong coupling regime occurs inevitably at vanishingly small carrier concentrations.

With the typical $\mu\approx0.1\,\mbox{eV}$ and minimal reasonable interlayer distance $D=50\,\mbox{\AA}$, at $\varepsilon=7$ (when weak coupling
approximation is still reliable) the expression (\ref{gap2}) gives $\Delta=4\times10^{-6}\,\mbox{eV}$, which is equivalent to the temperature of
$0.05\,\mbox{K}$. The maximal value of $\Delta$ can be achieved at strong coupling, when $\alpha\sim k_0D\sim1$. In this case $\Delta\sim\mu$, up to
hundreds of Kelvins.

Let now estimate the influence of disorder. In the case of conventional phonon-mediated superconductivity, a presence of magnetic impurities acts
destructively on BCS state because of a different scattering of electrons with opposite spins, which form Cooper pairs. In our case any impurity
localized in either graphene layer acts destructively on pair condensate, since it scatters only one pair constituent. The BCS-like state maintains
if the mean free path $\lambda$ exceeds the coherence length $l_\Delta=\hbar v_\mathrm{F}/\Delta$ (analogously to Ref.\,[25]). From the expression
for diffusive Boltzmann conductivity [26] we derive $\lambda\approx(\mu/ev_\mathrm{F})\mu_\mathrm{c}$, where $\mu_\mathrm{c}$ is the carrier mobility
in graphene. In dirty graphene samples $\mu_\mathrm{c}\approx1000\,\mbox{cm}^2/\mbox{V}\cdot\mbox{s}$ at room temperature [1] and the corresponding
mean free path is $\lambda\approx10\,\mbox{nm}$ at $\mu=0.1\,\mbox{eV}$. On the contrary, for clean graphene at the temperature of liquid helium
$\mu_\mathrm{c}\approx10^6\,\mbox{cm}^2/\mbox{V}\cdot\mbox{s}$, so $\lambda\approx10\,\mu\mbox{m}$. Comparing these quantities with estimations of
$l_\Delta$, we conclude that weak-coupling BCS-like state can be realized only in very clean graphene samples, whereas a strong-coupling state can
survive at rather strong disorder.

The onset of electron-hole pairing may be observed experimentally via drag effect peculiarities. It has been shown for CQW, that an occurrence of
pair condensation leads to a sharp increase of drag resistivity [27--29]. Another possibility is an observation of both stationary and non-stationary
Josephson-like effects (see, e.g., [30] and references therein). Also the condensation modifies an electromagnetic response of the system. In
particular, an application of in-plane magnetic field leads to a formation of persistent dipolar supercurrent, which can be detected directly
[14,~30,~31].

In CQW, the serious obstacle to an occurrence of a BCS-like electron-hole pairing is an anisotropy of the hole band [27,~32]. However in graphene the
particle-hole symmetry provides almost perfect matching in shape between electron and hole Fermi lines. In this connection it is interesting to
discuss a possible effect of the trigonal warping in graphene on BCS-like pairing in our system. The trigonal warping breaks the isotropy of Dirac
spectrum in graphene and causes a deviation of the Fermi line from a perfect circle towards a triangle-like shape [18]. The warping is negligible at
low carrier concentrations, but becomes considerable at large enough $\mu$. The triangle-like deviation has opposite orientations in two graphene
valleys and therefore breaks the valley symmetry, leading to the fixation of a condensate structure in valley space. However, in a case of small
trigonal warping a condensate with paired electrons and holes from different valleys will have slightly larger energy, than a condensate with
electrons and holed paired from same valleys. In such a situation, a two-gap state with new collective excitation modes can be formed, similarly to a
superconductor with overlapping bands [33].

Two spatially separated \emph{bilayer} graphene sheets can also be a candidate for a realization of the BCS-like pair
condensation. In a perpendicular electric field an electron spectrum in bilayer graphene acquires a tunable gap and
energy dispersion has a quite unusual ``Mexican hat" shape [34,~35]. After electrical doping a nontrivial Fermi surface
in the shape of a ring will be formed [36]. In principle, the BCS-like pairing between electron and hole Fermi-rings is
possible, leading to a formation of two energy gaps --- inside and outside of the ring.

In conclusion, we have analyzed a possibility of the BCS-like pairing between spatially separated electrons and holes
in two parallel graphene layers with negligible interlayer tunneling. At weak coupling the problem of pairing is rather
similar to that in CQW, except for several graphene-specific differences, i.e. Berry phase of electronic wave functions
and different screening properties. We have derived asymptotic expressions for the gap in the excitation spectrum at
various characteristics of the system and estimated its numerical value under reasonable conditions. Estimations of an
influence of disorder have been carried out. An appearance of the BCS-like electron-hole condensate can be observed
experimentally via Coulomb drag measurements, by studying a Josephson-like effect or by probing an electromagnetic
response of the system. The trigonal warping in the electron energy dispersion can lead to a formation of a two-gap
state.

There is no localized electron-hole pairs in graphene bilayer due to absence of a gap in the energy spectrum. Therefore
a behavior of the system on increase of the coupling strength is cardinally different from BCS-BEC crossover in CQW.
Both weak and strong coupling conditions can be achieved experimentally. If a dielectric constant of surrounding medium
is large enough, then the weak coupling regime sustains at arbitrarily small carrier concentrations, in contrast to a
situation in CQW.

Authors are obliged to participants of the Workshop on strongly correlated systems (Institute of High Pressure Physics)
for useful discussions of the results. Also the support from the Russian Foundation for Basic Research (Grant
06-02-81036-Bel-a) is acknowledged.

\end{document}